\documentclass[a4paper]{VisionStyle}
\usepackage{epsfig}
\usepackage{graphicx}

\begin{document}

\title{Timing with the EPIC pn Camera of XMM-Newton}

\author{M.\,Kuster\inst{1} \and E.\,Kendziorra\inst{1} \and
  S.\,Benlloch\inst{1} \and W.\,Becker\inst{2} \and U.\,Lammers\inst{3}
  \and G.\,Vacanti\inst{3,4} \and E.\,Serpell\inst{5}}

\institute{Institut f\"ur Astronomie und Astrophysik, Abt. Astronomie,
  Sand 1, D-72076 T\"ubingen, University of T\"ubingen,
  Germany 
  \and
  Max-Planck Institut f\"ur Extraterrestrische Physik -- MPE, 
  Giessenbachstr. 1, D-85784 Garching, Germany
  \and
  ESA Astrophysics Division (SA), Research and Scientific Support
  Department, ESTEC, Postbus 299, NL-2200 AG Noordwijk, The Netherlands
  \and
  Cosine Science \& Computing BV, Leiden, The Netherlands
  \and
  European Space Operations Center, ESOC, Robert-Bosch Str. 5, 64293
  Darmstadt, Germany
  }

\maketitle 
\begin{abstract}
  The EPIC pn CCD camera on board of \textsl{XMM-Newton} is designed to
  perform high throughput imaging and spectroscopy as well as high
  resolution timing observations in the energy range of 0.1--15\,keV. A
  temporal resolution of milliseconds or $\mu$-seconds, depending on the
  instrument mode and detector, is outstanding for CCD based X-ray cameras.
  In order to calibrate the different observing modes of the EPIC pn CCD,
  \textsl{XMM-Newton} observations of the pulsars PSR B1509-58, PSR
  B0540-69 and the Crab were performed during the calibration and
  performance verification phase. To determine the accuracy of the on board
  clock against Coordinated Universal Time (UTC), PSR B1509-58 was observed
  simultaneously with \textsl{XMM-Newton} and \textsl{RXTE} in addition.
  The paper summarizes the current status of the clock calibration.

\end{abstract}

\section{Introduction}

The \textsl{EPIC} instruments on board of {\sl XMM-Newton}, successfully
launched on 1999 December 10, are designed for imaging, high throughput
spectroscopy, and timing analysis. The EPIC pn camera offers six different
observation (readout) modes, four imaging modes, and two fast readout
modes.  The imaging modes are dedicated to high throughput imaging and
spectroscopy, while the fast readout modes are designed for high temporal
resolution down to 7.2~$\mu$s in combination with high spectral resolution.
For a detailed description of the EPIC pn readout modes, their
implementation, and characteristics see e.g.
\cite*{mkuster-WA2:kendziorra97a}, \cite*{mkuster-WA2:kendziorra99a},
\cite*{mkuster-WA2:kuster99a}, or \cite*{mkuster-WA2:ehle01a}.

This paper is structured as follows: In
section~\ref{mkuster-WA2_sec:readout} we give a brief description of the
fast readout modes available for the EPIC pn camera and their technical
implementation as far as these are of importance for the observer. In
section~\ref{mkuster-WA2_sec:obs-analysis} and
\ref{mkuster-WA2_sec:obs-results} we present preliminary results of the
relative time calibration based on observations of young millisecond
pulsars. In section~\ref{mkuster-WA2_sec:deadtime} we describe limiting
constraints an observer has to take into account when using the EPIC pn
camera for observations with high temporal resolution.

\section{Fast readout modes}\label{mkuster-WA2_sec:readout}
\begin{table}[b]
  \caption{Mode specific parameters for the EPIC pn camera. For the maximum
    possible time resolution the position of the source and the PSF has to
    be taken into account.}
  \label{mkuster-WA2_tab:tab1_periods}
  \begin{center}
    \leavevmode \footnotesize
    \begin{tabular}{lccr}
      Obs. Mode       &  Frame     & Life        &\multicolumn{1}{c}{Time} \\
                      &  Time [ms] & Time [ms]   &\multicolumn{1}{c}{resolution} \\\hline\hline
      Small Win.      &    5.672   &  4.028      &   5.67 ms       \\
      Timing          &    5.965   &  5.912      &  29.56 $\mu$s   \\
      Burst           &    4.345   &  0.126      &   7.2  $\mu$s   \\ \hline
    \end{tabular}
  \end{center}  
\end{table}
In addition to the imaging modes the design of the EPIC pn camera offers
two observational modes designated to temporal analysis of X-ray sources
with high time resolution, called Timing and Burst Mode. While Burst Mode
is designed for very bright sources up to 6.3 Crab only, Timing Mode is
usable for faint sources as well. In both modes spacial information in
y-direction is lost due to the continuous readout of the CCD. In contrast
to Timing Mode which uses the full size of the CCD, the last 20 lines of
the CCD are discarded in Burst Mode. To improve time resolution to the
maximum possible, the position of the source on the CCD has to be taken
into account. This is done within the \textsl{XMM Science Analysis System}
(\textsl{SAS}) for the nominal source position by default. For pointings
deviating from the nominal position, the observer has to specify the source
position on the CCD during the extraction process.

Beside high temporal resolution, both modes offer the full spectral
resolution of the EPIC pn camera. For a report on the status of energy
calibration see \cite*{mkuster-WA2:briel02a} or
\cite*{mkuster-WA2:kirsch02a} in this volume.

\section{Timing accuracy}\label{mkuster-WA2_sec:timing-accuracy}
All \textsl{XMM-Newton} data is tagged with a time stamp (on board time
OBT) from a 1 Hz on board clock provided by a temperature compensated
oscillator of the \textsl{XMM-Newton Command and Data Management Unit
  (CDMU)} (see \cite{mkuster-WA2:aranci98a}). The EPIC pn event analyzer
\textsl{EPEA} further uses this clock pulse as an input for its internal
clock which has an accuracy of 16 $\mu s$. All events handled by the
\textsl{EPEA} are tagged with this accuracy. A verification of the internal
timing of the \textsl{EPEA} was done during ground calibration at the
\textsl{PANTER} facility before launch (\cite{mkuster-WA2:kendziorra97a}).

On board time (OBT) is further converted to Coordinated Universal Time
(UTC) on ground while extracting data using information stored in a time
correlation file, delivered with each Observation Data File (ODF). This
file contains OBT versus UTC information. To transform event times, OBT is
correlated versus UTC using a polynomial fit. The user can influence the
accuracy of the fitting routine; for details see
\cite*{mkuster-WA2:lammers01a}.

During the extraction process with the \textsl{SAS}, the event times are
corrected according to the readout mode used during observation. In imaging
modes the photon arrival times are transformed to the center of integration
interval, for the best time resolution possible. In the fast readout modes
the time resolution can be improved by taking the position of the source
and the line in which the photon was detected into account. By default line
190 is used as source position, which is equal to the nominal pointing
position.

The last correction that has to be applied is a transformation to the solar
system barycentre. This is done via the \textsl{SAS} tool ``barycen'' which
transforms the given event times stored in the event file to Barycentric
Dynamical Time (TDB).

\section{Observations and data analysis}\label{mkuster-WA2_sec:obs-analysis}
\begin{figure}[t]
  \includegraphics[width=.99\columnwidth]{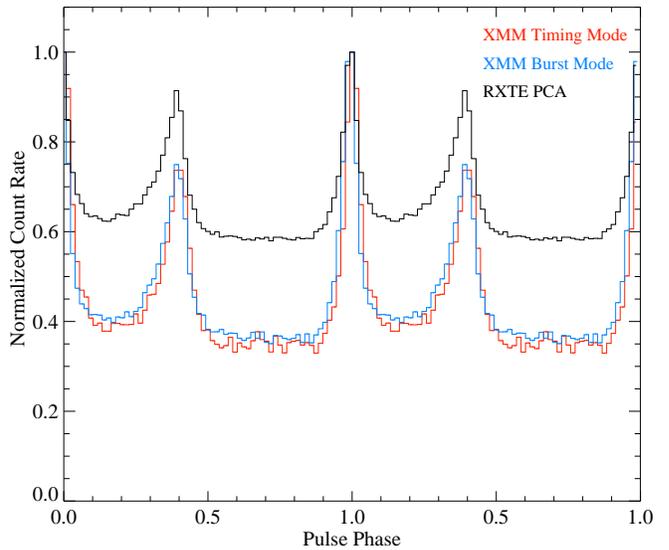}
  \caption{Pulse profiles of the Crab pulsar using \textsl{RXTE PCA} and
    \textsl{XMM-Newtons'} EPIC pn camera in Timing and Burst Mode. For
    \textsl{RXTE PCA} the pulse profile in the energy range 2.0--40\,keV is
    shown. For each pulse profile the maximum count rate is normalized to
    unity. Pulse phase zero is defined by the maximum count rate in each
    pulse, therefore absolute phase information is lost.}
  \label{mkuster-WA2_fig:Crab_pulse}
\end{figure}
\begin{figure}[t]
  \includegraphics[width=.99\columnwidth]{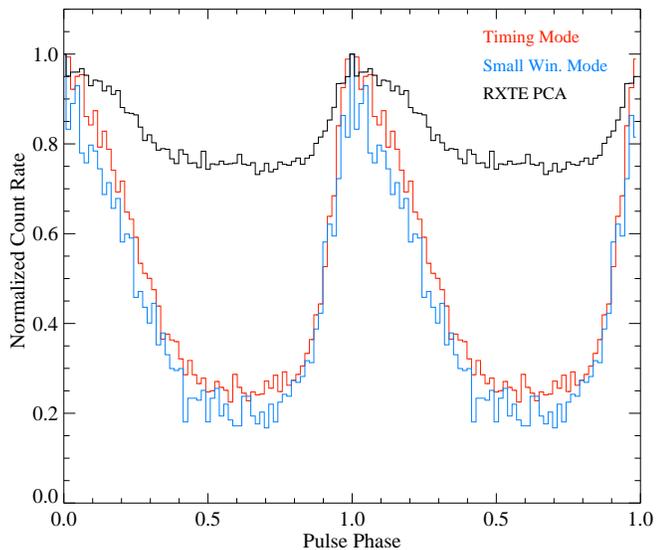}
  \caption{Pulse profiles of PSR B1509-58. The profiles are treated in the
    same way as the pulse profiles shown in
    Fig.~\ref{mkuster-WA2_fig:Crab_pulse}. Note that the small shift in
    phase between Timing and Burst Mode data is not a difference in
    absolute phase.}
  \label{mkuster-WA2_fig:PSR1509_pulse}
\end{figure}
During commissioning and performance verification phase of
\textsl{XMM-Newton} observations of young millisecond pulsars with the
purpose of relative and absolute clock calibration were made. In addition
we scheduled a simultaneous observation of PSR B1509-58 with \textsl{RXTE}
with purpose of absolute time calibration. A summary of all observational
data used for the analysis presented in this paper is given in
Tab.~\ref{mkuster-WA2_tab:observations}. For all Crab observations in
Timing Mode the electronic chopper was set to 25 to reduce telemetry rate.
All other observations were performed with electronic chopper set to 0.
\begin{table}[t]
  \caption{Observations used for time calibration.}
  \label{mkuster-WA2_tab:observations}
  \begin{center}
    \leavevmode
    \footnotesize
    \begin{tabular}{lcccccc}
      Object       &  
      Rev.         & 
      Mode &
      Chopper &
      Obs. Time \\
      &      &              &    & [ksec] \\ \hline\hline
      Crab     & 0056 & Timing       & 25 &   7.0 \\
      & 0234 & Burst        &  0 &  10.0 \\
      B1509-58 & 0137 & Timing       &  0 &   9.6 \\  
      & 0137 & Small Win.   &  0 &  11.4 \\ 
      B0540-69 & 0085 & Timing       &  0 &   7.0 \\
      & 0085 & Small Win.   &  0 &   7.0 \\\hline
    \end{tabular}
  \end{center}
\end{table}
Using the standard \textsl{SAS} procedure (\textsl{SAS} version 5.1.0) we
extracted event files and light curves for all observations, objects and
detector modes in the energy band 3.0--12.0\,keV. To reduce low energy
background to a minimum we ignored flux below 3\,keV. This is of importance
especially for observations during early commissioning phase, e.g. the Crab
observation in Rev.~56, when the detector setup was not yet optimized. For
observations in fast modes we used line~190 as position of the source in
RAWY direction for time correction. In addition we selected 4--5 col\-umns
around maximum intensity in RAWX direction as extraction region, to reduce
non pulsed flux from nebula emission or background. Further we transformed
photon arrival times to TDB, using the \textsl{SAS} tool ``barycen'' and
object coordinates given in Tab.~\ref{mkuster-WA2_tab:periods}.

From these light curves we derived pulse periods for each object and
observation mode using a $\chi^2$-maximization test. The resulting pulse
periods $P_{\rm found}$ are given in Tab.~\ref{mkuster-WA2_tab:periods}.
Using Jodrell Banks monthly data base for the Crab
(\cite{mkuster-WA2:Lyne01a}) and Princeton Pulsar Data base
(\cite{mkuster-WA2:taylor93a}), we extrapolated radio ephemeris $P_{\rm
  radio}$ for all sources using the center of observation time of the X-ray
observations as reference point. Subsequently we folded the X-ray light
curves with the respective pulse period to get pulse profiles shown in
Fig.~\ref{mkuster-WA2_fig:Crab_pulse} and
Fig.~\ref{mkuster-WA2_fig:PSR1509_pulse}. For all pulse profiles the
maximum flux is set to unity and pulse phase zero is defined by the maximum
flux. Therefore Fig.~\ref{mkuster-WA2_fig:Crab_pulse} and
Fig.~\ref{mkuster-WA2_fig:PSR1509_pulse} do not contain absolute phase
information.

\section{Observational results}\label{mkuster-WA2_sec:obs-results}
In Fig.~\ref{mkuster-WA2_fig:Crab_pulse} the resulting pulse profile for
the Crab is shown for Timing Mode, Burst Mode and \textsl{RXTE PCA}
observations.  The results for PSR B1509-58 in Timing Mode, Small Window
Mode and \textsl{RXTE PCA} are shown in
Fig.~\ref{mkuster-WA2_fig:PSR1509_pulse}.

As Fig.~\ref{mkuster-WA2_fig:Crab_pulse} and
Fig.~\ref{mkuster-WA2_fig:PSR1509_pulse} clearly demonstrate, we get
consistent pulse shape for all observational modes of \textsl{XMM-Newton}.
In addition the pulse profiles are well resolved and are in agreement with
those we get from \textsl{RXTE PCA} data. As described earlier we took
advantage of the imaging capability of \textsl{XMM-Newton} and selected
only the central emission region in RAWX direction during data extraction
in the fast readout modes. This reduces non pulsed flux from nebula
emission and thus we expect higher pulsed fraction for pulse profiles
derived from \textsl{XMM-Newton} data compared to \textsl{RXTE} data. The
larger amount of statistical uncertainties observed in \textsl{XMM-Newton}
data originates in lower effective observation time due to the electronic
chopper set during Timing Mode observation or due to low duty cycle in
Burst Mode.

\begin{figure}[t]
  \includegraphics[width=.95\columnwidth]{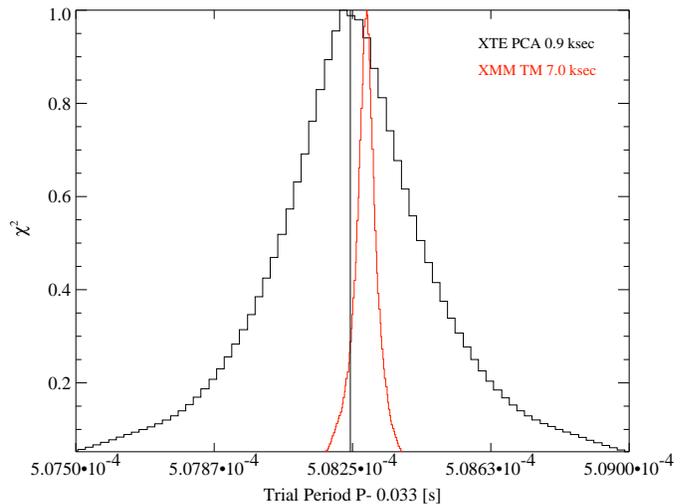}
  \caption{Resulting $\chi^2$-distribution of the period analysis of
    Crab data observed in Timing Mode in Rev.~0056 and \textsl{RXTE PCA}.
    The maximum of each distribution corresponds to the most significant
    pulse period. The vertical line marks the extrapolated radio period,
    the distribution drawn in red corresponds to \textsl{XMM-Newton} Timing
    Mode data. Note that the pulse period derived from XMM-Newton data is
    significantly longer compared to the radio period. The width of the
    $\chi^2$ distributions can not directly be used as error estimate for
    the pulse period.}
  \label{mkuster-WA2_fig:Crab_chisqr}
\end{figure}

As an example the resulting $\chi^2$ distribution of the period analysis of
the Crab observation of Rev.~0056 is shown in
Fig.~\ref{mkuster-WA2_fig:Crab_chisqr}. Within statistical errors the pulse
periods we get from \textsl{RXTE} data agree with the extrapolated radio
ephemeris marked as solid vertical line. For the \textsl{XMM-Newton} data
we get a deviation in pulse period of the order of $|\Delta P/P| \approx
10^{-6}$. This order of magnitude is consistent for all other objects and
readout modes, except Small Window mode which has a lower time resolution
(compare Tab.~\ref{mkuster-WA2_tab:periods}).  We calculated $\Delta P/P$
as

\begin{equation}
 \frac{\Delta P}{P}= \frac{(P_{\rm found}-P_{\rm radio})}{P_{\rm radio}}
\end{equation}

where $P_{\rm radio}$ is the extrapolated radio period. These discrepancies
are equivalent to a residual velocity component of 0.4--2.2 km/s which
might not be taken into account during the corrections applied to the
photon arrival times. To cross check our results with EPIC MOS, we analyzed
data of PSR B1509-58 observed in MOS timing mode during Rev.~0137, as well.
From this analysis we get pulse periods consistent with those derived from
EPIC pn data. The observed deviations are too large compared to the
specifications which would imply a $\Delta P/P \approx 10^{-8}$ and prevent
a absolute time cross calibration with \textsl{RXTE} for the time being.

To investigate the relative timing in more detail, additional calibration
observations of the Crab are scheduled. These observations will take place
at different orbital positions of \textsl{XMM-Newton} to be able to
separate residual orbital effects.
 
\begin{table*}[t]
  \caption{Pulse ephemeris resulting from temporal analysis. The expected
    pulse periods are extrapolated radio periods taken from Jodrell Bank
    monthly pulsar data base and the Princeton Pulsar data base. All radio
    pulse periods are extrapolated periods relative to the epoch
    given. $\Delta P/P$ is defined as $(P_{\rm found}-P_{\rm
    radio})/P_{\rm radio}$.} 
  \label{mkuster-WA2_tab:periods}
  \leavevmode
  \footnotesize
  \begin{tabular}{lccccrrr}
    Object       &  
    RA (J2000.0)  &       Dec (J2000.0) & Mode & Epoch      & 
    \multicolumn{1}{c}{$P_{\rm radio}$}  & 
    \multicolumn{1}{c}{$P_{\rm found}$}     &
    \multicolumn{1}{c}{$\Delta P/P$}        \\ 
    &            &           &      &  [MJD]     &                  
    \multicolumn{1}{c}{[ms]} &
    \multicolumn{1}{c}{[ms]} &              \\ \hline\hline
    Crab     & 05 34 31.973 & +22 00 52.061 & Timing   & 51632.8696 &
    33.508383 &  33.508424   & $ 1.3\times10^{-6}$  \\
    &              &               & Burst    & 51988.6595 & 
    33.521309 &  33.521238   & $-2.1\times10^{-6}$  \\
    B1509-58 & 15 13 55.617 & -59 08 08.872 & Timing   & 51794.3157 & 
    151.114141  & 151.113045  & $-7.3\times10^{-6}$ \\
    &              &               & Small & 51794.1681 & 
    151.114125  & 151.112350  & $-1.2\times10^{-5}$ \\
    B0540-69 & 05 40 11.049 & -69 19 55.188 & Timing   & 51691.5783 & 
    50.519308  & 50.519599   &$ 5.8\times10^{-6}$  \\
    &              &               & Small & 51691.0894 &  
    50.519288  &  50.519297  &$-1.9\times10^{-7}$  \\ \hline
  \end{tabular}
\end{table*}
\section{Dead time effects}\label{mkuster-WA2_sec:deadtime}
In order to quantify the efficiency of the time resolution with respect to
the dead time for each mode in the EPIC pn detector we have simulated
synthetic time series for Timing and Burst Mode. These light curves were
analyzed using standard time series analysis tools developed at our
institute.

The goal of our work was to determine the detection probability of a signal
in a power spectrum for the fast readout modes of \textsl{XMM-Newton}. In
all simulations our primary goal was to make use of the best possible time
resolution allowed in the respective mode. Note that some of the effects
described in the following sections, can partially be resolved by rebinning
the input light curve or the power spectrum (PSD), this however in any case
degrades the time or frequency resolution. A detailed description of the
simulations and the parameters used is given by
\cite*{mkuster-WA2:kuster99a}.

\subsection{Influence of readout sequence}
For our synthetic light curves we simulated random Poissonian noise with a
time resolution equal to the resolution of the observation mode, and with a
mean count rate that corresponds to the ``maximum'' count rate allowed for
the mode according to the telemetry and photon pile­up constraints (see
\cite{mkuster-WA2:ehle01a}). We subsequently folded the data with the
readout sequence (dead time function) for the respective mode, and computed
the power spectrum in Leahy normalization. We compared the resulting power
spectra with those expected for pure white noise.

\begin{figure*}[t]
  \begin{minipage}{0.48\textwidth}
    \includegraphics[width=0.97\columnwidth]{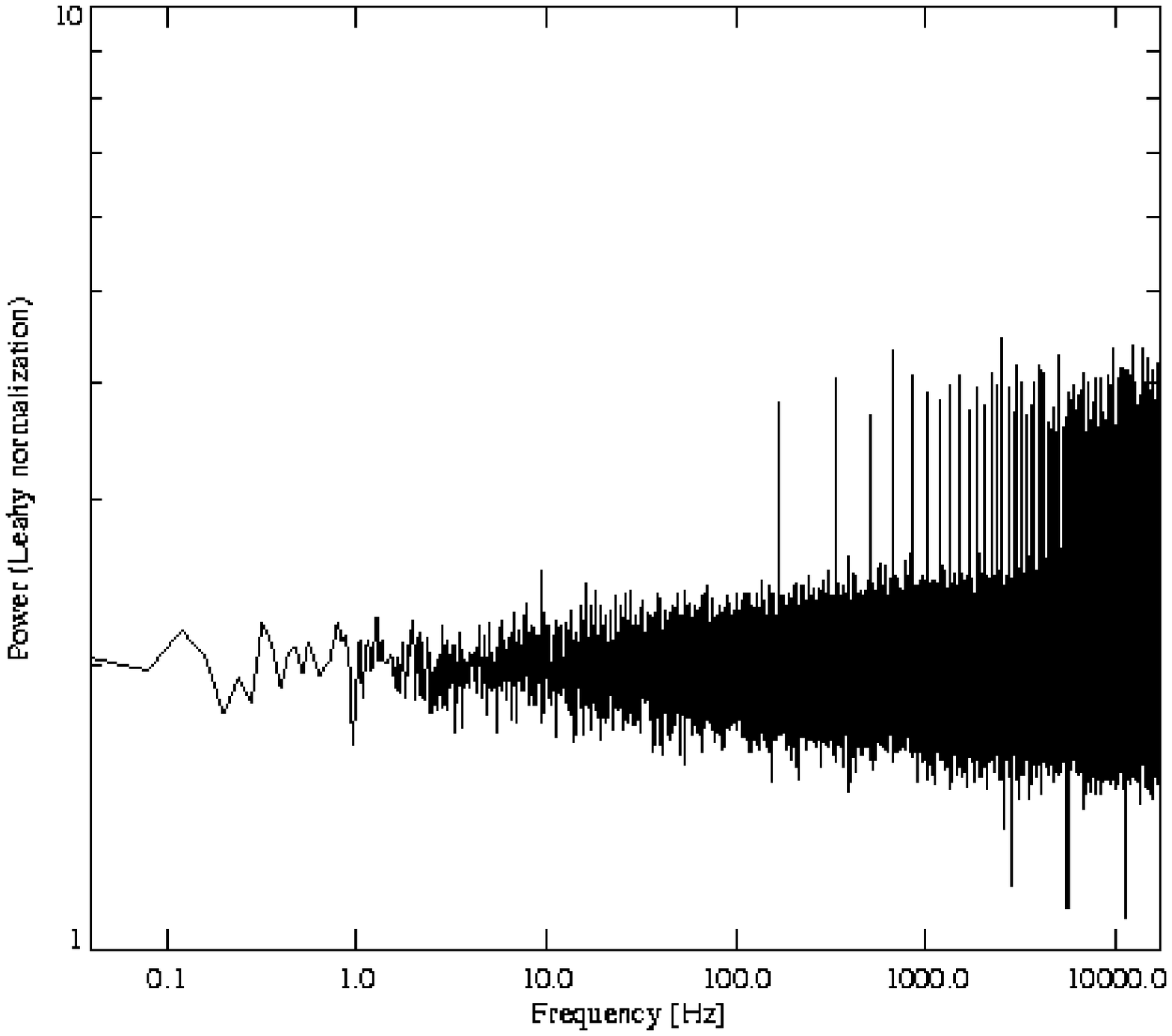}
  \end{minipage}
  \begin{minipage}{0.48\textwidth}
    \includegraphics[width=0.97\columnwidth]{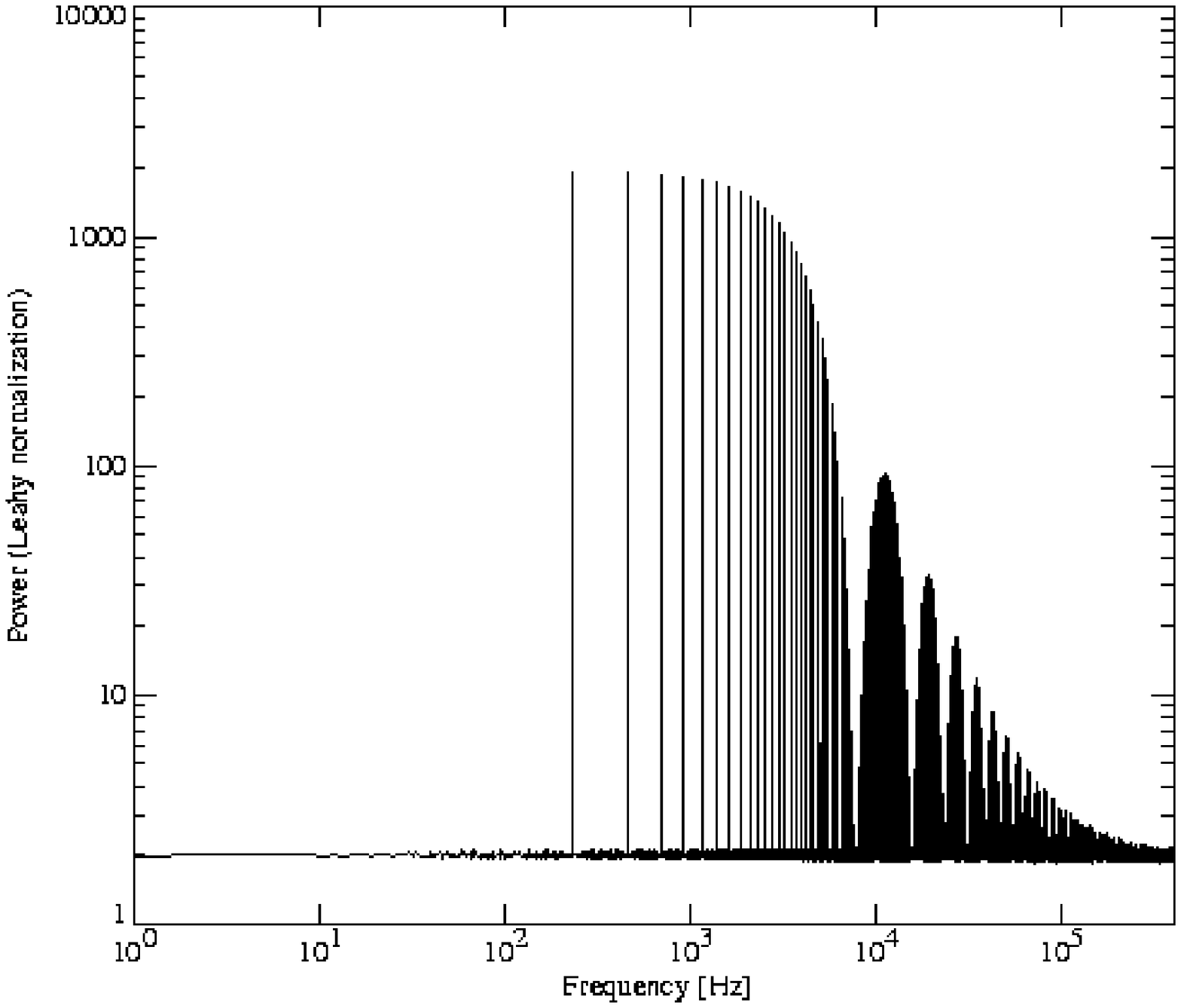}
  \end{minipage}
  \begin{minipage}{0.48\textwidth}
    \includegraphics[width=0.97\textwidth]{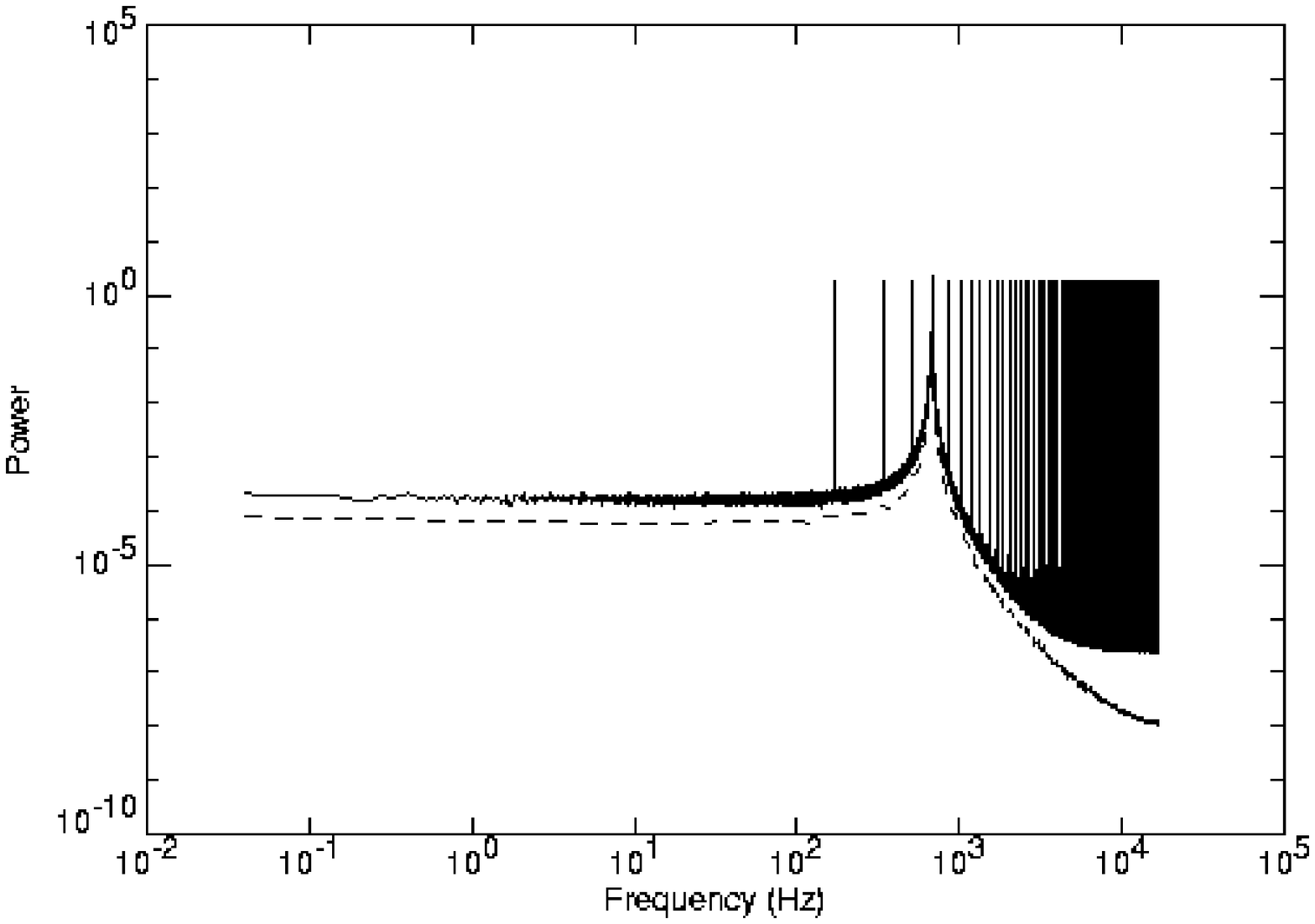}
  \end{minipage}
  \hspace*{0.5cm}
  \begin{minipage}{0.48\textwidth}
    \includegraphics[width=0.97\textwidth]{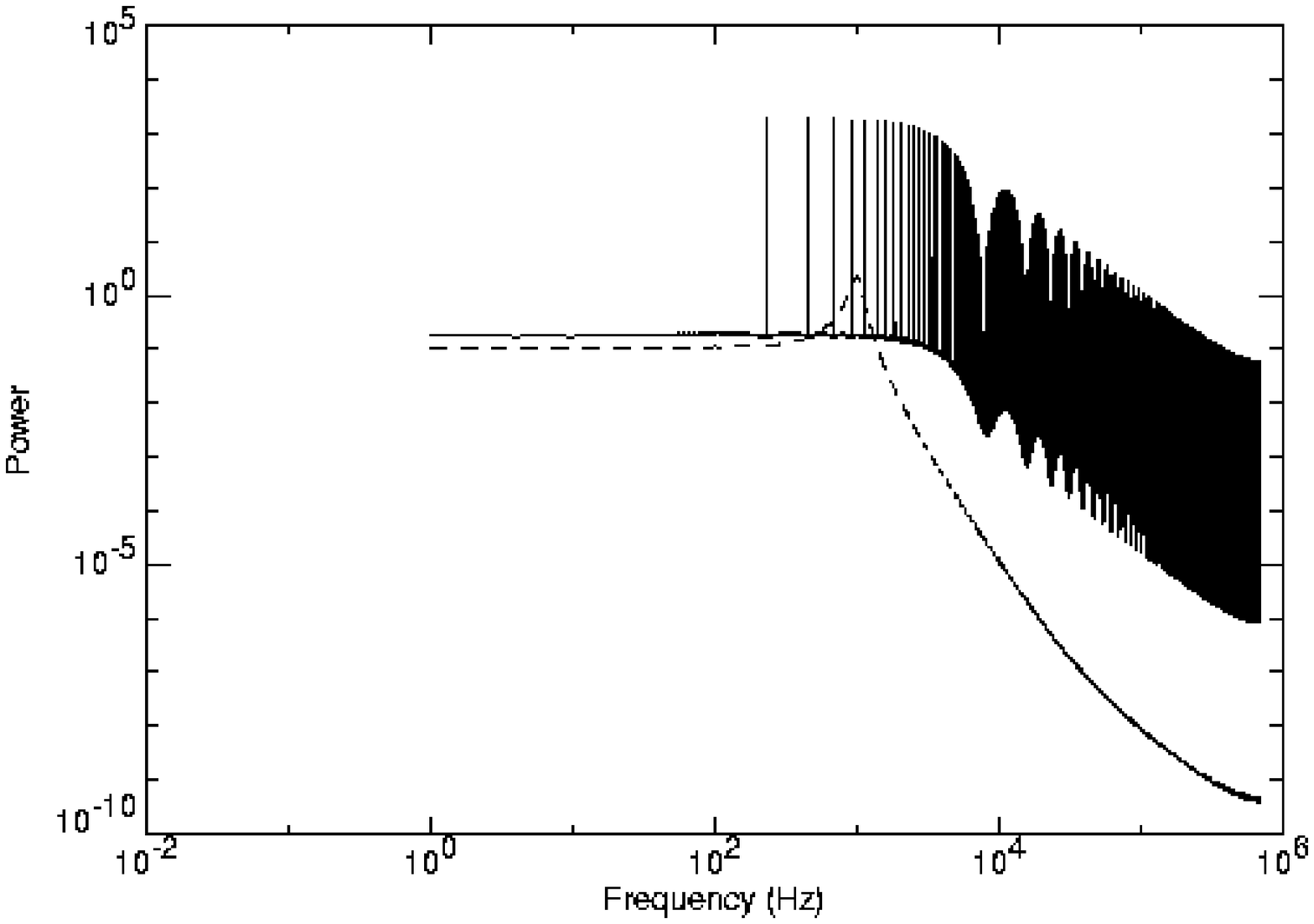}
  \end{minipage}
  \caption{Averaged PSDs out of 2500 power spectra assuming Poissonian
    noise data folded with the readout sequence of Timing and Burst Mode
    (top panel left and right). The same simulations assuming an additional
    signal of an AR[2] process representing a QPO (bottom panel left and
    right). The mean count rate for the simulations was set to 1500 cps,
    with a time resolution of 29.52 $\mu$s for Timing Mode and 60000 cps
    with a time resolution of 72 $\mu$s for Burst Mode. Note that the QPO
    can not be detected in Burst Mode.}
  \label{mkuster-WA2_fig:psd_tm_bm}
\end{figure*}

The results for Timing and Burst Mode show distorted features at large
frequencies which are produced by aliasing due to the dead time function of
the readout mode (see Fig.~\ref{mkuster-WA2_fig:psd_tm_bm}). The first peak
appearing in the power spectrum corresponds to the life/dead time window in
both cases: 5.91 ms (169.2 Hz) for the Timing Mode and 4.21 ms (237.5 Hz)
for the Burst Mode respectively. Consecutive peaks correspond to multiples
of these frequencies. Even with rebinning, for these high time resolution
modes the power spectrum is completely dominated by the windowing above
these ``critical'' frequencies, especially in Burst Mode.

In order to determine whether features in the PSD (e.g. QPOs) are
detectable above the ``critical'' frequencies, we analyzed the behavior of
the power spectrum in the presence of a variable source. To represent
realistic astronomical data, we have simulated a variable X-ray source
showing a quasi­periodic oscillation (QPO) at kHz frequencies on top of a
red noise spectrum and subsequently folded the resulting data with the
readout sequence. As the lower left PSD panel of
Fig.~\ref{mkuster-WA2_fig:psd_tm_bm} clearly demonstrates, for Timing Mode
the effect of the sampling rate has no influence on the behavior of the QPO
in the power spectrum, in the sense that the peak in the range of the
critical frequencies can be recognized between the well resolved peaks
produced by the sampling. On the other hand, for Burst Mode, the situation
is dramatically different. In this case the QPO cannot be detected above
the critical frequency (see lower right panel of
Fig.~\ref{mkuster-WA2_fig:psd_tm_bm}). The sampling rate effect completely
hides any sign of periodic (or quasi­periodic) oscillation in the data.

\subsection{Influence of detector configuration}
\begin{figure*}[t]
  \begin{minipage}{0.48\textwidth}
    \includegraphics[width=0.97\textwidth]{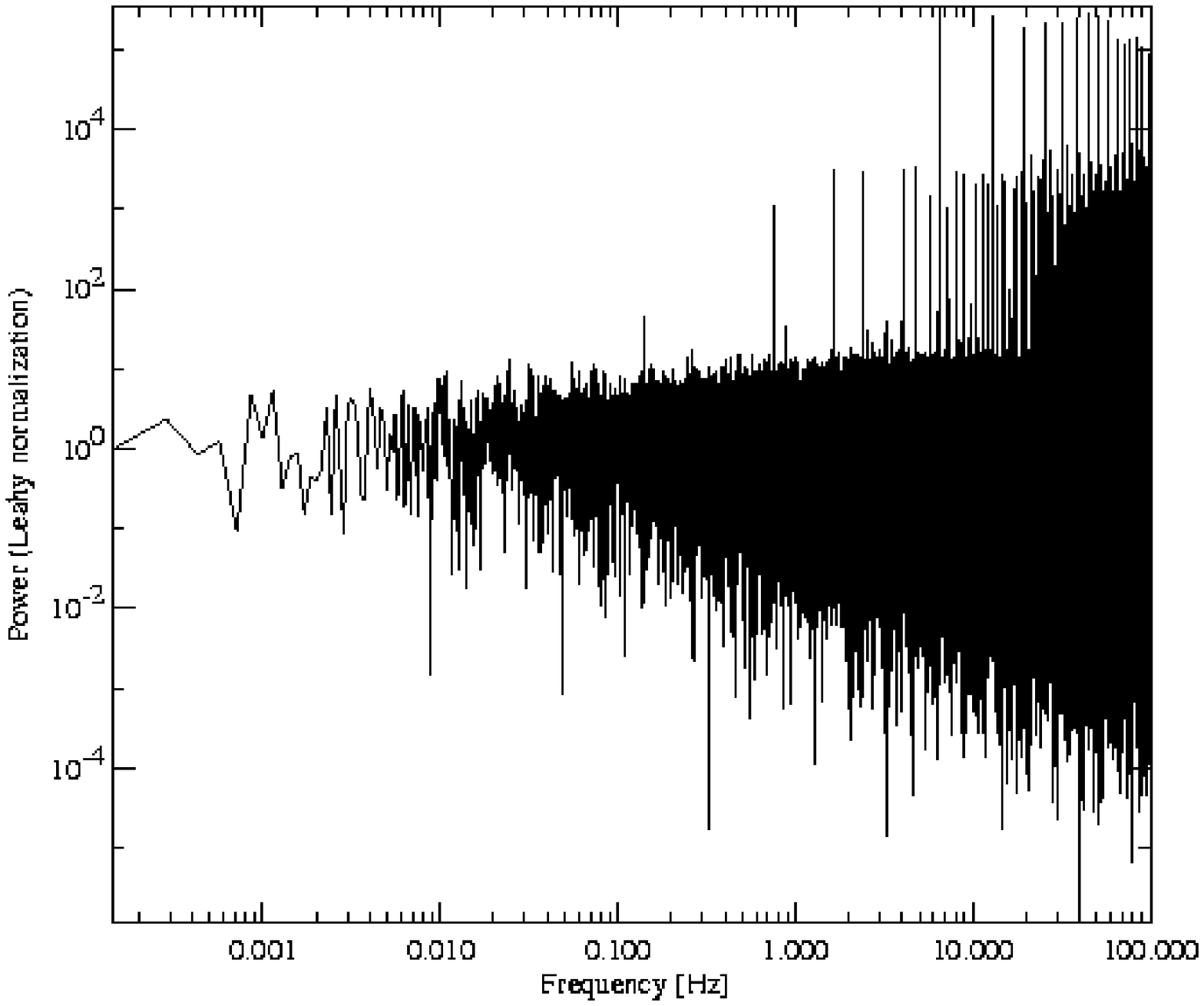}
  \end{minipage}
  \begin{minipage}{0.48\textwidth}
    \includegraphics[width=0.97\textwidth]{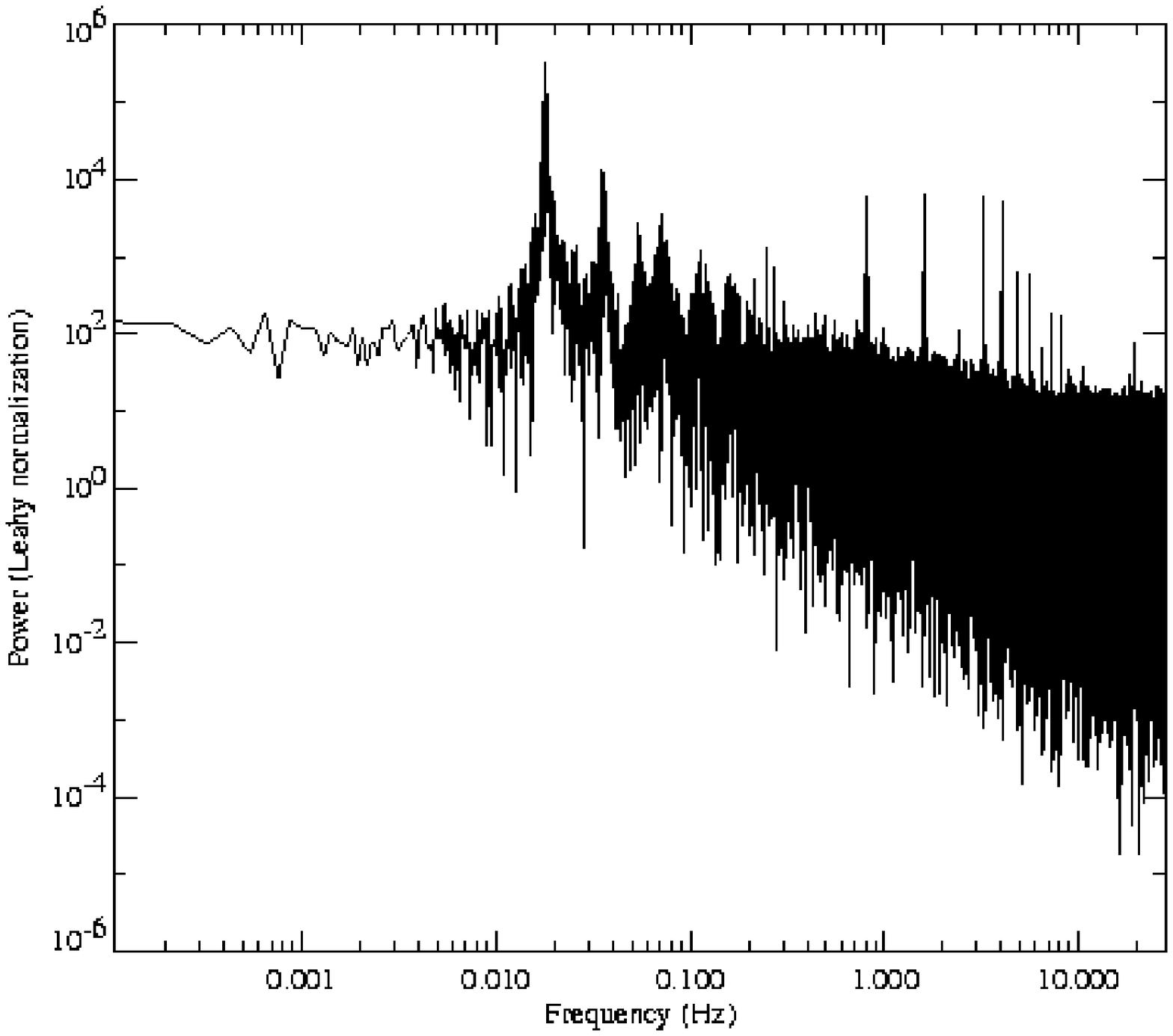}
  \end{minipage}
  \caption{Left: PSD of the Crab observation in Rev.~0056 during which the
    electronic chopper was set to 25. The PSD is totally dominated by noise
    above 10~Hz (compare Fig.~\ref{mkuster-WA2_fig:psd_tm_bm}). Right:
    During this pulsar observation the data handling unit periodically went
    to counting mode. The broad peaks below 0.2~Hz are caused by the
    counting mode. While the chopper increases noise at high frequencies,
    counting mode distorts the PSD at lower frequencies.}
  \label{mkuster-WA2_fig:counting_chopper}
\end{figure*}

In addition to the internal dead time of the pn-CCD due to the readout
sequence, the configuration of the detector can affect the time resolution
of an observation as well. In order to fulfill the telemetry constraints of
the \textsl{EPIC} Instruments especially for bright sources the observer
has the possibility to use the ``electronic chopper'' implemented in the
on board electronics of the EPIC pn camera (see
\cite{mkuster-WA2:ehle01a}).  For a chopper value of $N$, $N$ readout
frames of the CCD are discarded before processing each $(N+1)$~th frame.
Thus only the data of each $(N+1)$~th frame is added to the telemetry
stream. This adds an additional aliasing window function to the data
causing a distortion of the PSD. As an example the PSD of an Crab
observation in Timing Mode is shown in the left panel of
Fig.~\ref{mkuster-WA2_fig:counting_chopper}. During this observation the
electronic chopper was set to 25, thus reducing the effective life time by
the same factor. As a result the power spectrum is completely dominated by
noise above frequencies of 10~Hz.

Depending on count rate a similar effect is caused by counting mode. While
the electronic chopper adds a strict periodic dead time window, the effect
caused by counting mode generally is quasi-periodic. The observer should be
aware of these limiting constraints which are of importance expecially for
bright sources when planning an observation.
\section{Conclusions}\label{mkuster-WA2_sec:conclusion}
The fast modes of the \textsl{EPIC} cameras provide a time resolution which
is outstanding for X-ray CCDs. With this high time resolution it is
possible with X-ray CCD detectors to resolve pulsed emission even for fast
millisecond pulsars down to pulse periods of $\approx$~1 msec.

Our preliminary analysis of calibration observations of young millisecond
pulsars clearly demonstrates, that the internal timing of the EPIC pn and
EPIC MOS camera works as expected. However, currently the accuracy of
periods determined from \textsl{XMM-Newton} observations is only of the
order of $\Delta P/P \approx 10^{-6}$, deviating from an expected
uncertainty of $\Delta P/P \approx 10^{-8}$. This prevents a calibration of
the absolute time for the time being and needs further investigation.

Due to the readout principle of the detector the time resolution is limited
by aliasing frequencies caused by dead time effects. This is of importance
especially for bright sources and can partially be over come by rebinning
the light curve or the power spectrum. However, rebinning always diminishes
the effective time resolution as well. In addition the configuration of the
detector (electronic chopper or too high telemetry rates) can further
degrade the time resolution by additional periodic (el. chopper) or quasi
periodic (counting mode) dead time windows.

\begin{acknowledgements}
  We want to thank A. Rots who provided PSR B1509-58 data observed with
  \textsl{RXTE} for cross calibration purposes. This work is partially
  funded by ``Deutsches Zentrum f\"ur Luft und Raumfahrtangelegenheiten''
  (DLR) under grant 50 OX 0002.
  
  This work is based on observations obtained with \textsl{XMM-Newton}, an
  ESA science mission with instruments and contributions directly funded by
  ESA Member States.
\end{acknowledgements}

\appendix

\end{document}